\documentclass[aps,superscriptaddress,twoside,twocolumn,floatfix,a4paper,pra]{revtex4-2}
\newcommand{\comm}[1]{}

\usepackage{graphicx,epsfig}
\usepackage{color}
\usepackage[usenames,dvipsnames]{xcolor}
\usepackage[ocgcolorlinks,colorlinks=true,linkcolor=blue,citecolor=red]{hyperref}
\usepackage{amsmath,amssymb, amsthm, amsfonts}
\usepackage{dsfont} 
\usepackage{xspace}
\usepackage{xcolor}
\usepackage{changes}
\usepackage{verbatim}
\usepackage{mathtools}
\usepackage{amssymb}
\usepackage{comment}
\usepackage{soul}
\usepackage{siunitx}
\usepackage{dirtytalk}

\usepackage{booktabs,makecell,tabularx}
\usepackage{natbib}
\DeclareMathAlphabet{\mathbbold}{U}{bbold}{m}{n}

\newcommand{\one}{\mathbbold{1}}

\newcommand{\bb}[0]{\begin{eqnarray}}
\newcommand{\ee}[0]{\end{eqnarray}}
\newcommand{\ket}[1]{\vert #1\rangle}

\newcommand{\Cth}{$^{13}$C }
\newcommand{\cth}{{^{13}C}}
\newcommand{\Hone}{$^1$H }
\newcommand{\Fnine}{$^{19}$F }

\newcommand{\mc}[1]{{\color[rgb]{0,0,1}{{#1}}}}

\graphicspath{{Figures/}}

\begin{document}

\title{
Dynamical nuclear polarization for dissipation-induced entanglement in NV centers}

\author{Shishir Khandelwal}
\affiliation{D\'epartement de Physique Appliqu\'ee, Universit\'e de Gen\`eve, 1211 Gen\`eve, Switzerland} 

\author{Shashwat Kumar}
\affiliation{Institute of Physics, \'{E}cole Polytechnique F\'{e}d\'{e}rale de Lausanne (EPFL), Lausanne CH-1015, Switzerland}

\author{Nicolas Palazzo}
\affiliation{Institute of Physics, \'{E}cole Polytechnique F\'{e}d\'{e}rale de Lausanne (EPFL), Lausanne CH-1015, Switzerland}

\author{G\'eraldine Haack}
\affiliation{D\'epartement de Physique Appliqu\'ee, Universit\'e de Gen\`eve, 1211 Gen\`eve, Switzerland} 

\author{Mayeul Chipaux}
\thanks{Correspondence: mayeul.chipaux@epfl.ch}
\affiliation{Institute of Physics, \'{E}cole Polytechnique F\'{e}d\'{e}rale de Lausanne (EPFL), Lausanne CH-1015, Switzerland}

\vspace{10pt}

\date{\today}

\begin{abstract}
We propose a practical implementation of a two-qubit entanglement engine which denotes a scheme to generate quantum correlations through purely dissipative processes. On a diamond platform, the electron spin transitions of two Nitrogen-Vacancy (NV) centers play the role of artificial atoms (qubits), interacting through a dipole-dipole Hamiltonian.
%In this work, we put forward a practical implementation of a two-qubit entanglement engine, corresponding to an autonomous dissipation-induced entanglement generation scheme, utilizing Nitrogen-Vacancy (NV) centers on a diamond platform. We propose the electronic spin transitions of two NV centers to play the role of artificial atoms (qubits), interacting through a dipole-dipole Hamiltonian. 
The surrounding Carbon-13 nuclear spins act as spin baths playing the role of thermal reservoirs at well-defined temperatures and exchanging heat through the NV center qubits. In our scheme, a key challenge is therefore % environments with which the NV center qubits exchange heat. A key challenge in our scheme is
to create a temperature gradient between two spin baths surrounding each NV center, for which we propose the exploit the recent progresses in dynamical nuclear polarization, combined with microscopy superresolution methods. We discuss how these techniques should allow us to initialize such a long lasting out-of-equilibrium polarization situation between them, 
effectively leading to suitable conditions to run the entanglement engine successfully. 
Within a quantum master equation approach, we make theoretical predictions %Theoretical predictions are made within a quantum master equation approach, 
using state-of-the-art values for experimental parameters. We obtain promising values for the concurrence, reaching theoretical maxima.%, reaching maximal theoretical predictions. 
\end{abstract}

\maketitle

\section{Introduction}\label{sec:intro}
%\sk{SK: I think it is not useful to write about other methods of entanglement generation (that produce almost maximal entanglement) because our setup doesn't produce very high entanglement. Best to focus on something like }
%\\
%\mc{\textbf{\\ Motivations:}}

%can be achieved with very high fidelity through logical two-qubit gates, as demonstrated with fidelities exceeding $99\%$ within several experimental platforms, from superconducting~\cite{Bravyi2022} circuits to trapped ions~\cite{Bruzewicz2019}. Dissipation is unavoidable in quantum systems. Whether quantum correlations, in the form of entanglement for instance, can also be generated from purely dissipative processes, has been a challenging question in the past decades and triggered a series of theoretical proposals~\cite{BohrBrask2015a,Tavakoli2018,Tavakoli2020,Khandelwal2020,Aguilar2020,Das2022,Naseem2022}. 

\noindent The generation of quantum entanglement is a crucial task for quantum information processing.  Typically, this requires performing logical two-qubit unitary operations and limiting the quantum systems' interaction with their environment
%is typically achieved through logical two-qubit unitary operations which require limiting the quantum systems' interaction with their environment. For instance, the developed practical implementations,on superconducting circuits~\cite{Bravyi2022} and trapped ions~\cite{Bruzewicz2019} %, for example, strongly depend on the selected platform.%Practical 

These implementations require isolating the quantum systems from their environment, and platform-dependent techniques have been developed to achieve that, for superconducting circuits~\cite{Bravyi2022} and trapped ions~\cite{Bruzewicz2019}, for instance. 
However, avoiding all possible sorts of uncontrolled dissipation processes is impossible. In this context, natural questions
arises. Is it possible to exploit dissipation with thermal environments to create quantum resources? And how could this be implemented? %Expected to deeply contribute to the development of quantum technologies, this question
In recent years, those questions have triggered number of theoretical works,
leading to proposals for realizing thermal machines generating quantum correlations~\cite{BohrBrask2015a,Tavakoli2018,Tavakoli2020,Khandelwal2020,Aguilar2020,Das2022, Naseem2022}. It has also been demonstrated that the generated entanglement can be useful for performing non-classical operations~\cite{brask2022operational,Prech2023}. However, experimental demonstrations of these ideas are still lacking. In this work, we propose an experimental implementation of a two-qubit entanglement engine on a Nitrogen-Vacancy (NV)-center platform. 
\par
NV centers found in diamond are recognized as one of the most promising platforms for quantum technologies. As %was 
first demonstrated in 1997~\cite{Gruber1997}, they possess the features of an Optically Detected of Magnetic Resonance (ODMR) system~\cite{Suter2020ODMR}. This means that their electron-spin, coherently controlled with microwave, can be initialized and read-out with optical means. %, which means that their electron-spin can be coherently controlled by microwaves. Initialization and readout are performed 
In practice, this is performed with standard microwaves and visible range photo-luminescence microscopy equipment in the visible range~\cite{babashah2022optically}, %through standard photo-luminescence microscopy in the visible range \cite{babashah2022optically} 
or through Ground State Depletion (GSD) \cite{storterboom2021GsdNv} or STimulated Emission Depletion (STED) \cite{SilviaAdam2013StedNV} for better resolution. Combined with remarkable room-temperature quantum properties~\cite{Balasubramanian2009UltralongDiamond}, these qualities make NV centers particularly successful for quantum sensing applications \cite{malt, Barry2020NVSensitivityOptRev, Buergler2023}, while allowing for very versatile %under very different 
experimental conditions such as %; for example, 
high pressure \cite{Ho2021NvPressureReview} or%, 
high temperature \cite{Liu2019CoherentKelvin} conditions or % and 
within biological environments \cite{Chipaux2018NanodiamondsCells}. 
\par
NV centers also constitute a promising platform for quantum information processing \cite{Pezzagna2021QuantumDiamond,bradley2022robust}. Their fine and hyperfine structure with surrounding electronic or nuclear spins give them numerous controllable quantum degrees of freedom. Notably, NV centers can be exploited to control and entangle the dark electronic and nuclear spins of nearby P1 centers (corresponding to a nitrogen substitution)~\cite{Degen2021}. They are also often used as quantum bus to control multiple surrounding Carbon-13 (\Cth) nuclear spins operating as quantum memories~\cite{Bradley2019Ten-QubitRegister}. In parallel, Dynamical Nuclear Polarization (DNP) techniques \cite{broadway2018quantumHyperpol,Healey2021,Rizzato2022} have been developed to control the state of nuclear spins which are near the NV centers. These nuclear spins can be located in the bulk or at the surface of the diamond, and can correspond to different elements (\Cth, \Fnine or \Hone). These techniques were primarily developed towards hyperpolarized Nuclear Magnetic Resonance (NMR) applications \cite{Tetienne2021Prospectshyperpolarization}. From the perspective of open quantum systems, they constitute a path to realize effective spin baths with specific properties that can be controlled through the NV centers.

%Thanks to their numerous quantum degrees of freedom that can be controlled individually, diamond Nitrogen-Vacancy (NV) colored centers are among the most promising platforms for quantum information processing and quantum sensing for room temperature operation. In particular, both their triplet electron spin, and the nuclear spin of the associated Nitrogen, 14 or 15, can be used to as artificial atoms whose interaction with their surrounding Carbon 13 atoms can be control ed. Recently, great progress in dynamical nuclear polarization (DNP) evidenced that those Carbon 13 nuclear bath can be initialised and read by the NV center, with great potential towards magnetic resonance imaging and quantum information processing

%They allow for external control, tunability and measurements as the NV center's photo-luminescence ranges in the visible domain, leading to the development of various experimental techniques, Optical Detection of Magnetic Resonance (ODMR) \cite{Gruber1997} and standard photo-luminescence microscopy \cite{babashah2022optically}.

%\gh{gh: A short motivation why NV centers constitute promising platforms is NECESSARY. Please do it tomorrow morning - Mayeul and Shashwat and Nicolas. What should be present: NV centers, key properties that make them attractive for quantum technologies, what is DNP, as well as  microscopic super-resolution methods. say what is new (or not!) by combining these tools.}\\

In this work, we exploit recent developments in the field of NV centers to propose an experimental scheme for a thermal machine to produce entanglement between the electronic spins of two NV centers. Towards this goal, we put forward a novel application of DNP techniques to engineer two effective spin baths characterized by distinct polarizations. This also enters a broader field of applications that may require the control of the heat flow in NV-center technologies.

%The objective of this work is to propose a realistic implementation of this dynamics within a NV center platform for demonstrating dissipate-induced entanglement. We predict entanglement in the steady-state and transient regimes, exploiting the specificities of the NV centers.
%Depicted in Fig. \ref{fig:model}, the two qubits are constituted from the $\ket{m_S=0}$ to $\ket{m_S=-1}$ electron spin transitions of two diamond Nitrogen-Vacancy center. Positioned at a few nanometers from one another they interact through the dipole-dipole coupling. The two distinct thermal reservoirs are made of the \Cth nuclear spins directly surrounding each NV center.%\sk{, which can be realised through a gradient of polarization of the nuclear spins. 
%\mc{The NV-center relaxation is dominated by coupling to the \Cth reservoirs. MC: This will be say right after} 
%\mc{\textbf{\\ Outline}}

\par
The article is organized as follows. 
In Sec.~\ref{sec:model}, we briefly recall the basic ingredients \mc{of }%of the theoretical proposal for 
an entanglement engine to achieve dissipation-induced entanglement between two artificial atoms. 
In Sec. ~\ref{sec:Hsystem}, we propose the electronic spin of an NV center to play the role of an artificial atom, elaborating upon the various energy scales involved and the conditions to be satisfied to have steady-state entanglement between the electronic spins. In Sec.~\ref{sec:env}, we discuss how these electronic spins can be made to interact with their surrounding \Cth nuclear spin bath and show how a cross-relaxation regime may allow for purely dissipative dynamics. In Sec.~\ref{sec:tempbias}, we tackle the main challenges of this proposal, namely the realization of two thermal environments biased in temperature, combining dynamical nuclear polarization (DNP) techniques with microscopy superresolution methods. In Sec.~\ref{sec:ent}, we finally predict the presence of dissipation-induced entanglement between two NV centers using realistic theoretical predictions as well as state-of-the-art experimental parameters.%the realistic values derived in previous sections for the relevant parameters.  

\section{Theoretical model}
\label{sec:model}

A minimal model to investigate entanglement generation from dissipation is made of two interacting artificial atoms (qubits), each of them being tunnel-coupled to a thermal environment. These environments are assumed to be independent from one another, 
well-defined by their respective temperature and chemical potential through their Fermi distribution. When subject to a bias in temperature, a heat current flows between the two environments, through the system. If the qubits are interacting through a flip-flop type Hamiltonian (see below), this heat current has been demonstrated to sustain the presence of entanglement in the steady-state regime~\cite{BohrBrask2015a, Khandelwal2020, Heineken2021} and proposals were developed for semiconducting and superconducting platforms~\cite{Das2022, Naseem2022}. We now briefly recall the Hamiltonian of the two interacting qubits in this model, as well as the master equation that will be used to investigate the dynamics of this entanglement engine.

%Whereas each environment is characterized by a well-defined temperature at equilibrium, a temperature bias between them induces a heat current through the two qubits~\cite{Khandelwal2020}. Upon an adequate design of the interaction between the two qubits, which has been shown to be a flip-flop type interaction for two qubits, entanglement has been predicted both in the Non-Equilibrium Steady State (NESS), as well in the transient state (when the composite system has not yet reached its steady state)~\cite{BohrBrask2015a,Tavakoli2018,Heineken2021}.

The Hamiltonian of the two qubits is given by,
\bb
\label{eq:Hs}
H_{\text{\tiny S}} &=& 
\varepsilon_{\text{\tiny{L}}} \sigma_{\text{\tiny{L}}}^+\sigma_{\text{\tiny{L}}}^- \otimes \one_{\text{\tiny{R}}} +  \one_{\text{\tiny{L}}} \otimes \varepsilon_{\text{\tiny{R}}} \sigma_{\text{\tiny{R}}}^+\sigma_{\text{\tiny{R}}}^- \nonumber \\
&& + g (\sigma_{\text{\tiny{L}}}^+ \otimes \sigma_{\text{\tiny{R}}}^- + \sigma_{\text{\tiny{L}}}^- \otimes \sigma^+_{\text{\tiny{R}}})\,, 
\ee 
where $\varepsilon_\alpha\,\left(\alpha=\text{L, R}\right)$ are the energies of the left (L) and right (R) qubits and $g$ is the strength of the inter-qubit coupling. The raising and lowering operators for  qubit $\alpha$ are respectively $\sigma^+_\alpha$ and $\sigma^-_\alpha$. %and operators expressed as a tensor product of $2 \times 2$ matrices leave on the Hilbert space of the two qubits, spanned by computational states for instance $\{ \ket{00}, \ket{01}, \ket{10}, \ket{11} \}$. 
For a two-qubit device, it has been shown that a flip-flop type interaction Hamiltonian is suitable for entanglement generation \cite{BohrBrask2015a, Khandelwal2020}.  % $\sigma_\alpha^+$ and $\sigma_\alpha^-$. 
The qubits are distinctly coupled to two fermionic reservoirs (the choice is suitable for the NV-center platform, as we explain below). Assuming Markovian dynamics and weak system-reservoir couplings, the dissipative dynamics of the qubits can be described by a Lindblad master equation. Furthermore, if the interaction $g$, is small in comparison to $\varepsilon_\alpha$, %\sku{SKu: Should we define $\varepsilon_\alpha$?}
 the following local master equation is consistent~\cite{breuer2002theory},%\sk{[Refs]},

\begin{equation}\begin{aligned}\label{eq:lind}
\dot\rho(t) = &-i[H_{\text{\tiny S}},\rho(t)] \\
& + \sum_{\alpha\in\{\text L,\text R\}}\gamma^+_\alpha  \mathcal D\left[\sigma^+_\alpha \right]\rho(t)    +\gamma^-_\alpha \mathcal D\left[\sigma^-_\alpha  \right]\rho(t),
\end{aligned}
\end{equation}
with the dissipators defined as
$\mathcal{D}\left[A\right]\rho(t) \coloneqq A\rho(t)A^{\dagger} - \{A^\dagger A,\rho(t)\}/2$ and $\sigma^\pm_\alpha$ the raising and lowering operators. 
%\mcc{The Fermi distribution applies for distinguishable particles. For distinguishable ones (such as the \Cth) it should be the Boltzmann distribution.}
In case of fermionic baths, the Fermi distribution $n_F\left(\varepsilon_\alpha,T_\alpha\right)$ characterizing the environment of qubit $\alpha$ is evaluated at the qubit's energy $\varepsilon_\alpha$ and temperature $T_\alpha$. The total coupling rates are the product of the bare rate $\Gamma_\alpha$ (which depend on the microscopic details of the platform) and of the occupation probability of the environment \cite{Breuer2007},%, proportional to $n_F$ (occupied) and $1-n_F$ (empty) :
\begin{equation}
\begin{aligned}
\label{eq:DissiNF}
\gamma_\alpha^+&=\Gamma_\alpha\, n_F\left(\varepsilon_\alpha,T_\alpha\right) \\
\gamma^-_\alpha&=\Gamma_\alpha\left(1-n_F\left( \varepsilon_\alpha,T_\alpha\right)\right). 
\end{aligned}
\end{equation}
A temperature gradient between the two baths, $T_{\text{\tiny{L}}} \neq T_{\text{\tiny{R}}}$, induces a heat current flowing through the system. It was shown in \cite{BohrBrask2015a, Khandelwal2020} that, above a certain threshold for the heat current, entanglement between the two qubits is present in the steady state, but also in the transient regime. In the following sections, we discuss how to realize this model on a NV-center platform; first showing how the coupling between the NV-center qubits can be realized, and then showing how an effective temperature (polarization) gradient between the qubits can be created.

\section{Realization of the system Hamiltonian}
\label{sec:Hsystem}

% To start with, we discuss and characterize the realization of the two-NV center qubits Hamiltonian given in Eq.\eqref{eq:Hs}. 
%To enable the device to be further entangled they must hold sufficiently large large bare energies and be coupled through a flip-flop interaction.

\begin{figure*}[t]
    \centering
    \includegraphics[width=0.75\textwidth]{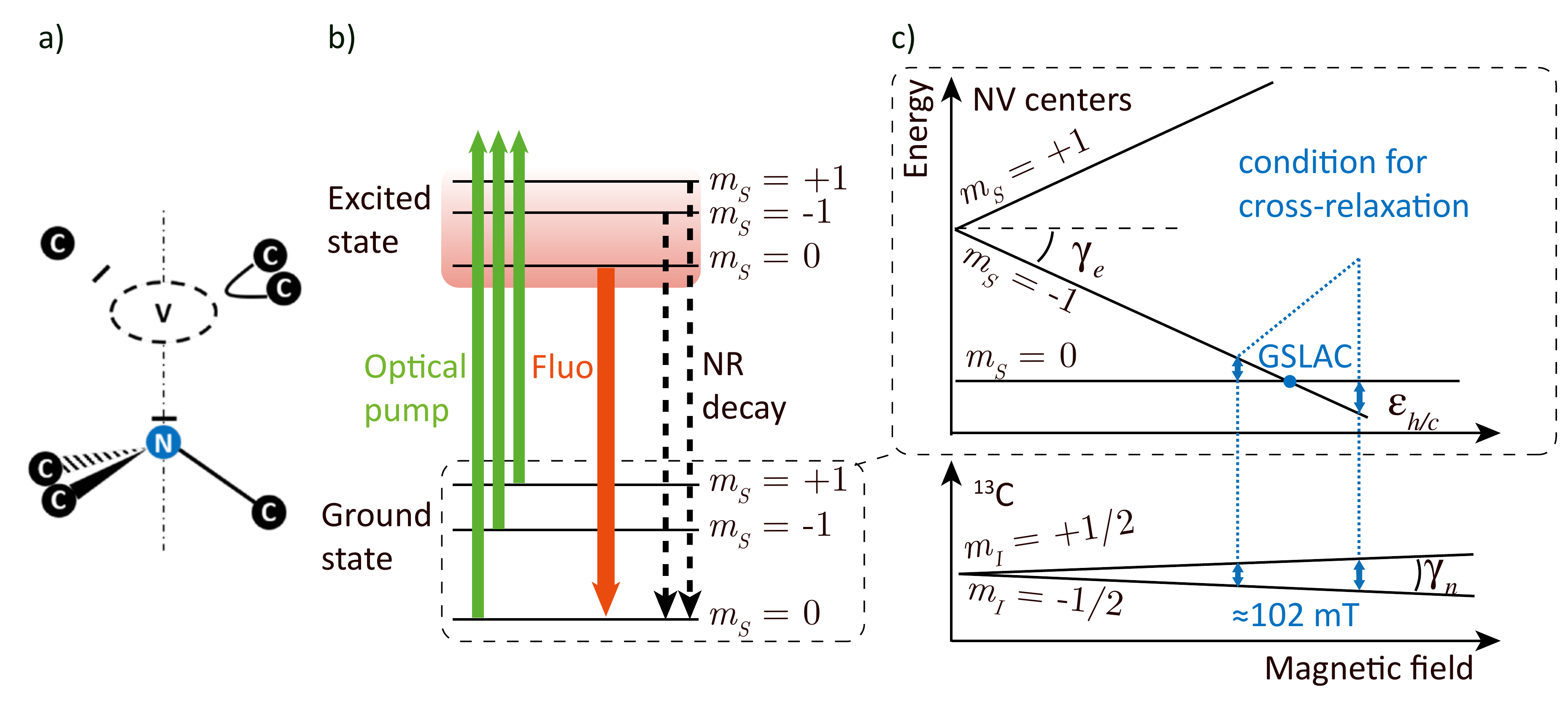}
    \caption{a) A representative sketch of a NV center in diamond; the Nitrogen atom ``N" next to the vacancy ``V", in a diamond lattice made of Carbon C. The electrons are shown in the Cram representation for 3 dimensional aspects. b) NV center spin-conservative electronic and optical transitions. The NV center is pumped by a green laser (typically at $\SI{532}{\,\nano\meter}$) and emits a red broadband photo-luminescence. A non-radiative decay path allows for spin non conservative transition and a polarization of the NV centers to its state of null spin. c) Energy diagram as a function of an aligned magnetic field. At certain magnetic field values, the transition of the NV center $m_S=0 \leftrightarrow m_S=-1$ may be at resonance with the one of the \Cth $-1/2 \leftrightarrow +1/2$. This corresponds to the cross-relaxation condition.}
    \label{fig:NVcenter}
\end{figure*}

\subsection{The diamond NV center}\label{subsec:NV}

The NV center's optical ground state is spin triplet. 
 The degeneracy between the non-zero states can be lifted in the presence of a magnetic through the Zeeman effect~\cite{Chipaux2015MagIma}. %The optical ground state of an NV center is a spin triplet, whose degeneracy can be lifted through the Zeeman effect in the presence of a magnetic field~\cite{Chipaux2015MagIma}. 
 Interestingly, the spin state can be initialized by an off-resonance optical green pump (e.g.  $520$ or $\SI{532}{\nano\meter}$) and read-out through its red photo-luminescence (from $600$ to $\SI{800}{\nano\meter}$). With $\vec{S} = \{ S_x, S_y, S_z\}$ the NV-center electronic spin and $m_S$ the associated spin quantum number, the NV center ground state Hamiltonian in a magnetic field $\vec{B}$ (Fig.~\ref{fig:NVcenter}) can be written as,
\begin{equation}
\label{eq:HNV}
H_{NV}=D S_z^2 + \gamma_e\vec{B}\cdot\vec{S} + H_{N} + H_{T},%
%,+ Q I_z^2 + \Gamma^{(i)}_{N}\textbf{B}\cdot\textbf{I} + \\  
% + &A_\parallel (S_xI_x + S_yI_y) + A_\perp S_zI_z,
\end{equation} 
%where $\vec{S} = \{ S_x, S_y, S_z\}$ is the NV-center electron spin. $\vec{B}$ is the external magnetic field, described below according to its longitudinal %parallel 
%$B_\parallel$ and transverse %orthogonal 
%$B_\perp$ components with respect to the the NV center quantization axis joining the nitrogen to the vacancy (Fig.~\ref{fig:NVcenter}). 
where the constant $D = 2\pi \times \SI{2.87}{\giga\hertz}$ is the Zero Field Splitting (ZFS) and $\gamma_e = g\mu_B/\hbar=2\pi\times\SI{28.0}{\giga\hertz\per\tesla}$ is the electronic-spin gyromagnetic ratio. $H_{N}$ and $H_{T}$ describe the hyperfine interaction of the NV center electronic spin $\vec{S}$ with the nitrogen atom ($^{14}N$ or $^{15}N$), and the surrounding \Cth (see Sec. \ref{sec:env}). 
%\gh{gh: here it would be interesting and useful to discuss the magnitude of these interactions, as we neglect $H_N$ right after.} 
%\mc{mc: better now?}
%\mc{$H_N$ just split the resonance into three (or two) lines spaced by 2 MHz (or 3 MHz) but the Zeeman effect remain the same.
%In the next paragraph we just neglect it so to recap the main effect of the Zeeman effect (and justify that we have to go at around 102 mT). The last paragraph of that subsection show that indeed, at GSLAC, $H_N$ can be negledted anyway as the nitrogen is is polarized to $M_I=0$ anyway. }\\
%\gh{The Hamiltonian $H_{NV}$ shows that} the NV center's energy can be \gh{controlled by} a magnetic field through the Zeeman effect. 
$B_\parallel$ and $B_\perp$ are respectively the longitudinal and the transverse components of the magnetic field with respect to the NV center quantization axis $\left(N-V\right)$ joining the nitrogen to the vacancy. %, when neglecting the hyperfine coupling with the Nitrogen atoms, 
The eigenenergies of the NV center corresponding to the spin states $\ket{m_s=\pm1}$ can be controlled through $B_\parallel$ according to the following equation \cite{Chen2013VectorMag}, up to first order in $B_\perp$ (see hereafter), $\varepsilon_\pm (B_\parallel) = D \pm \gamma_e B_\parallel$. In the following, we consider the energy transition from the ground state to the lowest excited state $\ket{m_s = -1}$ as the qubit transition, see top panel in Fig.1 c),
\bb
\label{eq:zeeman}
\varepsilon_{\text{\tiny{L,R}}} \equiv \varepsilon_- = D - \gamma_e B_\parallel\,.
\ee
To realize Hamiltonian of Eq.~\eqref{eq:Hs}, and the system evolution according to Eq.~\eqref{eq:lind}, it is essential that the two electronic spins are resonant not only with each other, but also with the nuclear spins of the surrounding \Cth. Under the assumption of a parallel magnetic field only, the energy of the nuclear spins of the \Cth is simply given by,
\begin{equation}
\label{eq:energy_nuclear}
\varepsilon_n = \gamma_n B_\parallel\,.
\end{equation}
The resonance condition imposes a condition on the modulus of the magnetic field to be close to the Ground State Level Anticrossing (GSLAC) condition,
\bb
&& \varepsilon_{\text{\tiny{L,R}}} = \varepsilon_n \label{eq:res}\\
&\Leftrightarrow& D - \gamma_e B_\parallel = \gamma_n B_\parallel \\
&\Leftrightarrow& B_\parallel \equiv B_{\text{\tiny{CR}}} = D/(\gamma_e + \gamma_n) \approx \SI{102}{\milli\tesla}\,. \label{eq:BCR}
\ee
For estimating $B_{\text{\tiny{CR}}}$, we have assumed $\gamma_n \ll \gamma_e$, valid for any nuclear spin with respect to any electronic spin. The label $CR$ for the magnetic field ensuring Eq.~\eqref{eq:res} refers to the energy-resonance condition, known as the cross-relaxation regime \cite{Broadway2016AnticrossingMagnetometry,Wood2017Microwave-freeScales}.

%\gh{gh:can we justify this?}\mc{Direct development of the Hamiltonian. I just added the reference \cite{Chen2013VectorMag}).}
%the orthogonal component of the magnetic field with respect to the $(N-V)$ quantization axis, and neglecting for now the hyperfine coupling with the nitrogen atom, the transition energies are:
%\bb
%\label{eq:zeeman}
%\varepsilon_\pm (B_\parallel) = D \pm \gamma_e B_\parallel\,.
%\ee 
%\mc{where the component in $B_\perp$ are at the second order only.}

%Now we discuss the other terms in Eq.~\eqref{eq:HNV}. The effect of $H_N$ is to split the two resonances into three (with $^{14}N$, triplet), or two (with $^{15}N$) respectively spaced by approximately $2$ or $\SI{3}{\mega\hertz}$ \cite{Felton2009Hyperfine}. Under the resonant condition, this contribution can be neglected. 

%As discussed in Sec.~\ref{sec:env}, they need to be configured in the cross-relaxation regime, where the nuclear spin energy matches that of the NV centers. Since the gyromagnetic ratio of any nuclear spin is orders of magnitude lower than electronic ones, $\gamma_n \ll \gamma_e$, this condition occurs near the Ground State Level Anticrossing (GSLAC). %, at a longitudinal magnetic field of $\approx \SI{102}{\milli\tesla}$ . This therefore imposes a specific value for both the applied magnetic field $B_{\text{\tiny{CR}}}$ and the qubit energies $\varepsilon_{L,R}$:
%\bb
%\label{eq:BCR}
%B_{\text{\tiny{CR}}}              &=&\frac{D}{\gamma_e+\gamma_n}\approx\SI{102}{\milli\tesla}\\
%    \label{eq:ELR}
%   \varepsilon_{L,R}   &=&\gamma_n\cdot B_{\text{\tiny{CR}}}=\SI{1.1}{\mega\hertz}.
%\ee 

The energy-resonance regime must be reached while preserving the NV centers' quantization axes and their ODMR properties \cite{Tetienne2012Transverse}. This imposes that the applied magnetic field must be well aligned with the quantization axis $\left(N-V\right)$ of the two qubits, hence $B_\perp\approx0$ as in Eq.~\eqref{eq:zeeman}. %\gh{gh: Mayeul: is this correct? before we wrote that we expand up to first order in $B_perp$: compatible?}.  %the NV quantization $\left(N-V\right)$ axis (i.e., on which $B_\perp\approx0$). %This ensures that the NV center quantization axis, and therefore its ODMR properties, are preserved \cite{Tetienne2012Transverse}. 
Subsequently, this also imposes that the two NV centers are along the same crystallographic orientation, aligned with the magnetic field. In Fig. \ref{fig:NVcenter}, we provide a schematic view of a NV center, its electronic and optical transitions, as well as the energy diagram with the key conditions to satisfy with the applied magnetic field. 

We now discus the additional terms $H_N$ and $H_T$ in Eq.~\eqref{eq:HNV}. Near GSLAC, $H_{N}$ plays an important role which depends on the type 
of nitrogen isotope used in the experiment \cite{AuzinshHuijie2019GSLAC,Ivady2021GSLAC}. As shown in Ref. \cite{Broadway2016AnticrossingMagnetometry}, the energy transition between states $\ket{m_s=0}$ and $\ket{m_s=-1}$ of a NV center comprising of the spin doublet $^{15}N$ isotope cannot be lowered below $\SI{2}{\mega\hertz}$ and reach the \Cth nuclear transition. This makes $^{15}$NV unsuitable for our proposal. Instead, the nuclear triplet $^{14}$N, naturally much more abundant ($>\SI{99.5}{\percent}$) than the other isotopes, satisfies this condition and will be considered here. As discussed in detail in Sec. \ref{sec:env}, this isotope has already been demonstrated to reach the cross-relaxation regime with \Cth \cite{broadway2018quantumHyperpol} and other nuclear spins of higher gyromagnetic ratio \cite{Wood2017Microwave-freeScales}. Finally, with $I_n$ the nuclear spin operator and $m_I$ its associated spin quantum number, close to GSLAC, the $^{14}$NV center nuclear spin is naturally polarized into its non-interacting $\ket{m_I=0}$ state in a durable manner \cite{Ivady2021GSLAC,Dreau2013SingleShot}, which explains the absence of role played by $H_T$ in this regime.

%In the following, we will therefore omit $H_N$, and consider that each NV center can be modeled \gh{as} a simple two-level system between $\ket{m_S=0}$ and $\ket{m_S=-1}$ with a transition energy labeled $\varepsilon_{L,R}$. % given by $\varepsilon_{L,R}$ in Eq.~\eqref{eq:ELR}.

\subsection{Inter-qubit dipole-dipole interaction}
\label{subsec:Hdd}
The electronic spins of two NV centers interact naturally through a dipole-dipole interaction, whose strength depends on the electromagnetic gyromagnetic ratio $\gamma_e$ of each NV center and on the relative distance between them characterized by the vector $\vec{r} = r \hat{e}_{\text{\tiny{R}}}$ below. Within the secular approximation (valid at high magnetic field), the dipole-dipole interaction Hamiltonian takes the form \cite{slichter2013principles,MagResApp2014},%respect to their decoherence can be neglected with respect to mutuatheir interaction. sk{SK: Can this be rephrased? maybe saying that the distance is short, so the interaction is large compared to the decoherence.}%NV centers to form a pair.
%Given that the two NV centers need to be quantized along the same axis, (see discussion above)
%where $\theta_{ij}$ is the angle between the quantization axis and the vector $\Vec{r}_{ij}$ characterizing the relative position of each NV center with each other.
%\begin{equation}\begin{aligned}
%\label{eq:dip-dip}
 %   H_{dd}&=\frac{\mu_0\gamma_e\%  \left(\Vec{S}_{\text{\tiny{L}}}\otimes \Vec{S}_{\text{\tiny{R}}} - (\Vec{S}_{\text{\tiny{L}}}
%    \cdot \hat{e}_{\text{\tiny{R}}})(\Vec{S}_{\text{\tiny{R}}}\cdot \hat{e}_{\text{\tiny{R}}})\right)
%\end{aligned}\end{equation}
%with its development in terms of the dipolar alphabet:
\bb%\begin{equation}\begin{aligned}
\label{eq:dip-dip2}   
H_{LR}&=&\frac{\mu_0 \gamma_e\gamma_e \hbar}{4\pi r^3} \big( 1 - 3 \cos^2 \theta \big) \nonumber \\ 
&&\left( S_{\text{\tiny{L}}}^z S_{\text{\tiny{R}}}^z - \frac{1}{4} (S_{\text{\tiny{L}}}^+ S_{\text{\tiny{R}}}^- + S_{\text{\tiny{R}}}^+ S_{\text{\tiny{L}}}^-)  \right)\,,
\ee%\end{aligned}
%\end{equation} 
where the first two terms correspond to the terms $A$ and $B$ of the dipolar alphabet of the dipole interaction Hamiltonian. Let us note that, for clarity, we have omitted the tensor product in the above equation compared to Eq.~\eqref{eq:Hs}. In the literature, the case of two off-resonant nuclear spins is typically discussed such that 
the second term is neglected. In this proposal, we consider the regime of resonant nuclear spins for which the Hamiltonian above should be considered \cite{Dikarov2016SolidFlipFlop}. In particular, one notes that the second interaction term realizes the flip-flop interaction required by Eq.~\eqref{eq:Hs} with the coupling constant $g$ set by,
\begin{eqnarray}
\label{eq:g1}
\vert g \vert  = \frac{\mu_0 \gamma_e^2 \hbar }{4 \pi r^3} \frac{1-3\cos^2\theta}{4}.
%    g = \frac{\mu_0 h\gamma_e^2}{4\pi r_{LR}^3}\times \frac{1-3 \cos^2⁡\theta_{LR}}{2}
\end{eqnarray} 
The angle $\theta$ denotes the angle between the orientation NV-center axis and the direction of $B_\parallel$ ($B_{\text{\tiny{CR}}}$). The contribution of the $S^z_{\text{\tiny{L}}} S^z_{\text{\tiny{R}}}$ term only plays a role for a certain class of initial states, those who are not described by a $X$-shaped density operator. The predictions we make in the following sections are done considering relevant experimental initial states, which are all of a $X$-shape. For other initial states, this term will induce additional non-zero off-diagonal terms in the transient regime, but will not affect the steady state. We can therefore neglect this term without loss of generality for this proposal.

For a successful realization of a NV-based entanglement engine, interaction strength $g$ must  within the right energy range with respect to bare energies of the NV centers and relaxation rates $\Gamma_{\text{\tiny{CR}}}$ with the surrounding \Cth (discussed just below), $ \Gamma_{\text{\tiny{CR}}} \leq g \ll \varepsilon$. The interaction strength $g$ directly depends on the distance between the two NV centers. In recent years, stochastic implementation of pairs \cite{Yamamoto2013StronglyImplantation,Jakobi2016EfficientDiamond,Zhao-Jun2016GenerationImplantation} and triplets \cite{Haruyama2019TripleImplantation} of NV centers has been achieved in experiments. The implantation energy determines the final depth and straggling \cite{Pezzagna2010NVCreatio} of the nitrogen atoms in the diamond, and therefore the expected distance that separates them. % the expected distance that separates the NV centers. 
Based on magnetic spectroscopy measurements, coupling strengths ranging from a few $\SI{}{\kilo\hertz}$ as limited by decoherence, up to the record of $\SI{3.9}{\mega\hertz}$ %as 
reported in Ref. \cite{Jakobi2016EfficientDiamond} have been demonstrated, falling well into the above range. The main experimental values of the parameters of the model that correspond to the discussions in the present and following sections are summarized in %App. \ref{app:a} 
Table \ref{tab:1}. \par
We now discuss environment engineering and relaxation energy scales.

\section{System-environment interaction}
\label{sec:env}

In this section, we first discuss the role of different environments for an NV center and then propose a scheme to engineer an out-of-equilibrium situation between two NV centers, which is required for entanglement generation.

%thermal environments inducing a dissipative dynamics for the interacting NV centers playing the role of qubits, and then propose a scheme to mimic a temperature gradient between two distinct environments, which the goal of inducing a thermal heat current through the system. % here the coupling of each NV qubit to their \Cth environment and show how their quantum dynamics can be dominated by the desired dissipative process. 

%\subsection{Dipole interaction}
\subsection{Dipolar interaction}
%\subsection{Environment}
\label{subsec:env}

A key advantage of the NV-center platform is the excellent decoupling of NV centers from lattice phonons~\cite{Gugler2018AbDiamond}, with a room temperature spin-lattice relaxation rate %$\Gamma^{SL}$ 
as low as a few hundreds of Hertz. As a consequence, %relaxation time and
quantum coherence properties of NV centers are mainly limited by surrounding magnetic impurities  \Cth in our case\cite{Barry2020NVSensitivityOptRev,Balasubramanian2009UltralongDiamond,Mizuochi2009,Bar-Gill2013Solid-stateSecond}. %\gh{gh for Mayeul: is the previous statement correct??}. 
%It is the dipole-dipole Hamiltonian that describes the interactions between the electronic spin of a NV center, and the nuclear spins of the $j$-th \Cth nuclear spin \cite{Barry2020NVSensitivityOptRev,Balasubramanian2009UltralongDiamond,Mizuochi2009,Bar-Gill2013Solid-stateSecond}. 

The interactions between a NV center's electronic spin $\alpha$, and the $j$-th \Cth's nuclear spin is again described by the dipole-dipole Hamiltonian.
Within the secular approximation, under resonant conditions discussed in Eqs.~\eqref{eq:res} and \eqref{eq:BCR}, the electronic spin $\alpha$ (with gyromagnetic ratio $\gamma_e$) interacts with $N$ \Cth nuclear spins (with gyromagnetic ratio $\gamma_n$) with the Hamiltonian $H_{T,\alpha}$,

\bb
\label{eq:Ht2}
H_{T,\alpha} &=& \sum_{j=1}^N \frac{\mu_0  \gamma_e\gamma_n \hbar}{4 \pi {r_{\alpha j}}^3} \big( 1-3 \cos^2{\theta_{\alpha j}} \big) \nonumber  \\
&&\left( S_\alpha^z I_j^z - \frac{1}{4} (S_\alpha^+ I_j^- + I_j^+ S_\alpha^-)  \right)\,.
\ee
Here, $\vec{S}_{\text{\tiny{L,R}}}$ and $\vec{I}_j$ denote the spin and nuclear operators respectively. The term $ S_\alpha^z I_j^z$ will lead to pure dephasing via the rate $\Gamma_{2,\varphi}$ of the electronic spin of the NV centers, while the tunneling (flip-flop) term is at the origin of dissipation (decoherence rate $\Gamma_{2,\text{\tiny{CR}}}$ and relaxation rate $\Gamma_{1 \alpha}$). In the following subsections, we discuss in further detail these rates and predicted values in the context of our proposal.

\begin{figure*}%[H]
\includegraphics[width=0.7\textwidth]{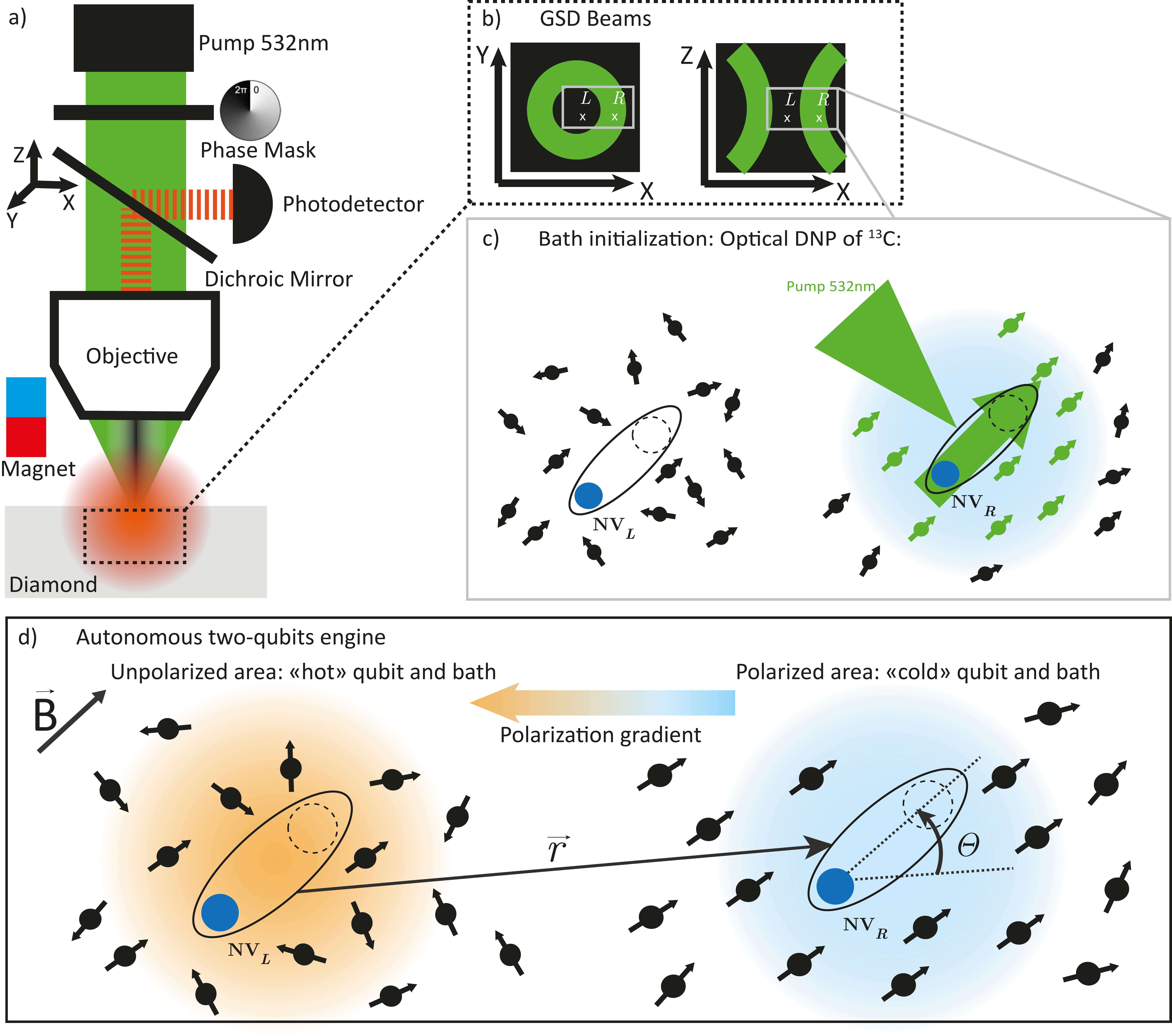}%[scale = 0.1]{spindif.png}
\centering
\caption{a) Experimental setup with ground state depletion capabilities. b) Pumping beam profile at the system level. c) Dynamical Nuclear Polarization on one NV center, cooling the surrounding \Cth spin bath. d) Configuration of the autonomous two-qubit entanglement engine, with the optical pumping beam off.}

\label{fig:model}
\end{figure*} 

\subsection{Cross-relaxation rate}

The effect of the \Cth nuclei on the relaxation of the NV center $\alpha$ is known to be determined by the product of the root mean squared value of the  transverse magnetic field noise, $B_{\text{rms},\alpha}$, and its spectral density $S_\alpha(\varepsilon)$ \cite{slichter2013principles,Sigaeva2023NOstart},
\bb
\label{eq:R1}
    \Gamma_{1, \alpha}(\varepsilon) &=& 3 \gamma_e^2(B_{\text{rms}, \alpha})^2 S_\alpha(\varepsilon).%, \,\, i\in\{L,R\}.
\ee
In full generality, the spin-lattice relaxation rate $\Gamma_{\text{\tiny{SL}}} $ needs to be added to this expression, if non-negligible. Considering the contributions of all \Cth nuclei to be independent from each other, $(B_{\text{rms}, \alpha})^2$ can be obtained by summing up all the contributions squared \cite{slichter2013principles},

%\mcc{This step, from the sum of hamiltonians, to $(B_{rms}^{\alpha})^2 $ and $ S_\alpha(\varepsilon)$ is exactly equivalent than applying the Lindblad formalism: 
%\begin{itemize}
%    \item markovian approximation
%    \item We do not consider the back action from the qubit to the \Cth but only the effect of the \Cth on the spin via $\left<{B_{\perp}^{\alpha}}^2\right>$ 
%    \item We consider each \Cth as independent from one another
%    \item Going from a discreet sum to an integral is also equivalent to a infinite bath ...    
%\end{itemize}
%}

\bb
\label{eq:brms}
   (B_{\text{rms}, \alpha})^2  = \sum_{j}{\left(\frac{\mu_0\gamma_n\hbar}{4\pi}\right)}^2C_S\frac{2+3\sin^2{\theta_{\alpha j}}}{{r_{\alpha j}}^6},
\ee 
where $C_S=1/4$ is linked to the multiplicity of the reservoir spins. % and $r_j$ is the position of the $j^{th}$ \Cth with respect to considered qubit $L$ or $R$.
The number of nuclear spins contributing to the sum over $j$ can be estimated through the volume of influence \cite{Terblanche2013CDefects}. Nuclear spins that are closer than a minimal radius around the NV center $R_{m}\approx\SI{0.2}{\nano\meter}$, as well as the ones beyond a maximal radius $R_{M}\approx\SI{1.33}{\nano\meter}$ are excluded. In a diamond with a natural abundance of \Cth of $n_C = \SI{1.1}{\percent}$, we estimate the sum over $j$ to comprise approximately 20 spins, which will define the environment of the electronic spin of NV center $\alpha$. Assuming $r_{\alpha j} = r_\alpha$ and the same orientation of the nuclear spins with respect to the quantization axis of spin $\alpha$, $\theta_{\alpha j} = \theta_\alpha$, Eq.~\eqref{eq:brms} can then be written as,

%\mc{Noting $\vec{r}_\alpha=r_\alpha\vec{u}_\alpha$ the position with respect to the qubit $\alpha$ and $\theta_\alpha$ it angle with respect to $\left(N-V\right)$ we approximate Eq.~\eqref{eq:R1} with a continuous description integrating the \Cth contribution from $R_m$ to infinity \cite{Tetienne2012}}
\begin{equation}
\begin{aligned}
\label{eq:brms2}
(B_{\text{rms}, \alpha})^2  &= C_S\left(\frac{\mu_0\gamma_n\hbar}{4\pi}\right)^2  \iiint_{r>R_{m}}\frac{2+3\sin^2{\theta_{\alpha}}}{r^6} n_C dV, \\
  &=  C_S\left(\frac{\mu_0\gamma_n\hbar}{4\pi}\right)^2 \frac{7\pi}{2 {R_{m}}^3}n_C,
\end{aligned}
\end{equation}where $n_C$ is the average \Cth density, or its natural abundance. 
%This is overestimated since the probability to find a \Cth that close to the NV center is very small while through in this continuous description its takes an important effect. This can be corrected by utilizing the probability to find each next \Cth at a distance $r$. \cite{Hall2014CentralSpin}. 
%With this we can give a better numerical estimation of $\sqrt(\gamma_e^2B_{rms}^{\perp\alpha}^2)$ which is then modulated by $S_(\varepsilon_{L,R})$
The spectral density $S_\alpha(\varepsilon)$ is approximated by a Lorentzian, centered on the nuclear spin energy $\varepsilon_n$ (see Eq.~\eqref{eq:energy_nuclear}), and whose linewidth is limited by the NV center decoherence $\Gamma_{2,\alpha}$ \cite{Wood2016WideBand, Hall2014CentralSpin},
\bb
S_\alpha(\varepsilon) = \frac{1}{\Gamma_{2,\alpha}}\times\frac{1}{1+{(\varepsilon-\varepsilon_n)}^2/\Gamma_{2,\alpha}^2}.
\ee
In the most common off-resonance case ($\varepsilon_\alpha \gg \varepsilon_n$), the spectral density vanishes such that the nuclear spin induced relaxation remains small with respect to the spin lattice one. 

Instead, in the cross relaxation condition which applies here ($\varepsilon_{\text{\tiny{L,R}}} = \varepsilon_n$,), the maximal relaxation rate, hereby called cross-relaxation rate, has the form:
\bb
\label{eq:gamma_Rel}
\Gamma^{\alpha}_{1,\text{\tiny{CR}}} = \frac{3 \gamma_e^2(B_{\text{rms}}^{\alpha})^2}{\Gamma_{2,\alpha}}.
\ee

%In contrast, far from resonance, $\varepsilon_\alpha \gg \varepsilon_n$, the spectral density vanishes, and dissipation through cross-relaxation vanishes. In this off-resonant regime, the spin-lattice relaxation contribution becomes dominant (order of a few 100 Hz \gh{gh for Mayeul: can you check the units?}[\sk{sk: Is this kHz or Hz?}]) and can not be neglected anymore \cite{Gugler2018AbDiamond}.

It has been measured up to $\Gamma_{1, \alpha} = \SI{250}{\kilo\hertz}$ in a diamond with a natural concentration of \Cth \cite{broadway2018quantumHyperpol}, three orders of magnitude larger than the spin-lattice relaxation rate $\SI{200}{\hertz}$ in the same sample. 

For our proposal, considering the natural abundance $n_C$, the cross-relaxation bare rate $\Gamma_\alpha$ in Eq.~\eqref{eq:DissiNF} will be taken to be,
\begin{equation}
\label{eq:gcrcth}
\begin{aligned}
\Gamma_1 \equiv \Gamma_{1,\alpha}\left(\%(\cth)\right) = \frac{\%(\cth)}{\SI{1.1}{\percent}}\times
\SI{250}{\,\kilo\hertz}\,.
\end{aligned}
\end{equation} 

In a similar sample diamond, the decoherence rate is measured to be of the order of $\SI{1.5}{\kilo\hertz}$ \cite{Mizuochi2009}, with a linear increase with $n_C$:%. For our predictions in the last section, we therefore consider,
\begin{equation}
\begin{aligned}
\label{eq:g2cth}
\Gamma_{2,\alpha} \left(\%(\cth)\right)=\frac{\%(\cth)}{\SI{1.1}{\percent}}\times\SI{1.5}{\kilo\hertz},
\end{aligned}
\end{equation}
This remains two order of magnitude lower than the relaxation rate. In the cross relaxation regime, the dynamics of the NV centers' electron spins can therefore be modeled by the purely dissipative cross-relaxation process, for which Eq.~\eqref{eq:lind} should be valid.
In the following, we
consider the same bare rates for the electronic spins coupled to their respective nuclear spin baths, $\Gamma_{\text{\tiny{L}}} = \Gamma_{\text{\tiny{R}}} = \Gamma_{1}$ given by Eq.~\eqref{eq:gcrcth}.

\section{Thermally biased environments}
\label{sec:tempbias}

In this section, we explain how to engineer two artificial thermal environments made of \Cth nuclear spins, biased in temperature. In an NV-center platform, this is made possible owing to the NV centers being naturally decoupled from the phonons \cite{Gugler2018AbDiamond}, allowing us to not be subject to the excellent thermal conduction properties of the diamond. Here, we realize an effective temperature difference between two ensembles of nuclear spins through Dynamical Nuclear Polarization (DNP). We predict this technique to enable a long lasting out-of-equilibrium situation within the context of this proposal. The main steps are shown in Fig. \ref{fig:model} and discussed below. 

\subsection{Dynamical Nuclear Polarization}
\label{subsec:DNP}

For the nuclear spins with energy $\varepsilon_n$ within the volume of influence of each NV center (at a position $\vec{r}$ and time $t$), it is possible to estimate their polarization $p = p_{\downarrow} - p_{\uparrow}$ by assuming a Boltzmann distribution for each of the nuclear spin state $p_{\downarrow} = 1/(e^{-\varepsilon_n/(k_B T)}+1)$ and $p_{\uparrow} = 1 - p_{\downarrow}$,
\begin{equation}
    p = p_\downarrow - p_\uparrow = \frac
    {1-e^{
    -{\varepsilon_n/k_b T}
    }}
    {1+e^{-{\varepsilon_n}/{k_b T}}}
    %\approx \frac{\gamma_n B_{\text{\tiny{CR}}}}{2 k_b T}\,,
\end{equation}
%\gh{gh for Mayeul: expansion valid only for small energies. Do we really need to make the expansion? Can we not just keep the exact ratio and estimate it to $10^{-7}$?}. 
%with $p_\uparrow + p_\downarrow =1$. Here we have assumed $B_\parallel = B_{\text{\tiny{CR}}}$ to be in the cross-relaxation regime as discussed in the previous section. At $T=0$ K, $p\downarrow = 1, p_\uparrow =0$, whereas at infinite temperature, $p_\downarrow = p_uparrow = 1/2$.
At room temperature, with $\varepsilon_n = 1.1$ MHz and $B_{\text{\tiny{CR}}} = 102$ mT, we obtain $p \approx 10^{-7}$, corresponding to a situation without any effective temperature gradient.\par
%Assuming a continuous description, we note $p_{\downarrow}\rt$ and $p_{\uparrow}\rt$ the probability to find, at time $t$, each spin at the position $r$ in its $\ket{+\frac{1}{2}}$-state and $\ket{-\frac{1}{2}}$-state respectively $\left(p_{\downarrow}\rt+p_{\uparrow}=1\right)$.
%\gh{the nuclear states have never been introduced before! Do we need them?}\\ We then define the polarization as 
%\bb\label{eq:polar}
 %   p\rt=p_{\downarrow}\rt-p_{\uparrow}\rt\,,
%\ee
%which is $0$ at infinite temperature and ${1}$ at $\SI{0}{\,\kelvin}$. %It may also equals $-1$ for up polarized spin. According to the Boltzmann distribution $\frac{p_{\uparrow}}{p_{\downarrow}} = \exp{\left(-\varepsilon_n/k_b T\right)}$, the polarization is given by,
%at a temperature $T$ and under a magnetic field $B$ the nuclear spin thermal polarization, e.g. before DNP is applied equals 
%\bb
%   p = \frac {1-e^{{\varepsilon_n/k_b T}}}{1+e^{-{\varepsilon_n}/{k_b T}}}\approx \frac{\gamma_n B_\parallel}{k_b T}.
%\ee
Several methods have been proposed and experimentally demonstrated to lower or increase the polarization of nuclear spins by transferring the polarization of an electronic spin in their vicinity. These techniques are known as Dynamical Nuclear Polarization (DNP). With NV centers, a common technique is the Hartmann-Hahn (HH) protocol \cite{Hartmann1962}, which drives the electron and nuclear spins in a synchronous manner. It has been adapted to the NOVEL sequence \cite{Shagieva2018NvNovel,Rizzato2022}, using spin locking to adjust the Rabi frequency in the rotating frame to the nuclear Larmor frequency, or to the more recent PulsePol schemes \cite{Schwartz2018PulsPol,Healey2021}. In the cross-relaxation regime, which is relevant for this proposal, no radio frequency fine tuning is necessary; DNP is achieved through optical pumping of the NV center. Recently, a cross-relaxation induced polarization (CRIP) protocol has demonstrated a polarization as high as $99 \%$ for \Cth and Fl \cite{broadway2018quantumHyperpol}.  DNP uses the NV center as a source of polarization via the spin selective non-radiative decay path, which allows the preparation of a nuclear target spin into its $\ket{m_S=0}$ state \cite{Gruber1997}. While polarization of a single nuclear spin through DNP is estimated to happen in tens of ns, it was shown experimentally that polarization of nuclear spins within the volume of influence of a NV center takes of the order of a few $\mu$s \cite{Shagieva2018NvNovel}.\par
DNP protocols are therefore foreseen to be suitable to cool down nuclear spins within the volume of influence of a given NV center, hence realizing an effective thermal bath characterized by an effective temperature that is lower than room temperature. A remaining challenge for our proposal is the ability to select the NV center for applying DNP, i.e. to achieve a selective cooling. We discuss this below.

\subsection{Selective cooling}
\label{subsec:GSD}

To achieve selective cooling of nuclear spins surrounding a given NV center, we envision two possible experimental schemes.

%At this stage, it remains to assign a temperature to a given bath, in other words, to perform a selective cooling (polarization) of only one of the two thermal reservoirs. To achieve that, we propose to use one NV center to polarize the surrounding \Cth spin bath while the other is untouched. The corresponding experimental scheme is illustrated in Fig.~\ref{fig:model}.
%\par
%Two strategies can be foreseen.

The first involves distinguishing the two thermal baths by their electron paramagnetic spectrum \cite{bersin2019individualQubitControl}. A \Cth nuclear spin close to one NV can shift its resonance such that a DNP protocol could target it specifically. This however would involve having qubits of different energies which is less favorable to generate entanglement; see \cite{Khandelwal2020} for theoretical predictions. Another approach, shown in Fig.~\ref{fig:model}, involves using a superresolution microscopy method such as Ground State Depletion (GSD) \cite{storterboom2021GsdNv} or STimulated Emission Depletion (STED) \cite{SilviaAdam2013StedNV}. % to target one of the two NV centers at once. 
In the case of GSD, the pumping beam (green) is patterned to have a central dark spot on the focus plane (see Fig.~\ref{fig:model}b)). The NV center within the dark spot is therefore not optically pumped while the other is. In the case of STED, a similarly patterned depletion beam \cite{SilviaAdam2013StedNV} is added to a conventional green pump. The NV center placed in the dark center of that red beam is pumped normally, while the excited state of the other NV center is depleted such that its spin-non conservative decay path is impeached. In either case, only one of the two NV centers can be regularly pumped and act as a local source of polarization for a DNP protocol. \par
To summarize, in our proposal, the hot bath is realized by considering the nuclear spins surrounding the left NV center (for instance), the effective hot temperature being the room temperature corresponding to a polarization close to 0. %(see discussion above).
The cold bath will be realized through DNP and selective cooling as discussed above and shown in Fig.~\ref{fig:model}. The predicted effective cold temperature is estimated through the achieved polarization in the range $97 \% - 100\%$. With these experimental achievements, the rates entering the Lindblad master equation Eq.~\eqref{eq:lind} for the left (hot) and right (cold) baths can be identified as:
\bb
&&\gamma_{\text{\tiny{L}}}^+ = \gamma_{\text{\tiny{L}}}^- = \frac{1}{2} \, \Gamma_{1}\,, \\
&&\gamma_{\text{\tiny{R}}}^- = \frac{1 +p_{\text{\tiny{R}}}}{2} \, \Gamma_{1}\,=\left(1-n_F\left( \varepsilon,T_{\text{\tiny{R}}}\right)\right)\Gamma_1, \\
&&\gamma_{\text{\tiny{R}}}^+ = \frac{1 -p_{\text{\tiny{R}}}}{2} \, \Gamma_{1}=n_F\left( \varepsilon,T_{\text{\tiny{R}}}\right)\Gamma_1\,,
\ee
with $p_{\text{\tiny{R}}}\in\left[0.97, 1\right]$ 
%\gh{gh for Shishir: it is an interval. I took it from your plot. Would you like to formulate it differently?}\sk{[is this an interval or a set?]}
and $\Gamma_{1}$ given by Eq.~\eqref{eq:gcrcth}. Before discussing the numerical predictions for this NV-center based entanglement engine, we comment on spin diffusion and spin-lattice relaxation that would, in general, affect  DNP protocols.

\begin{figure*}
    \centering\hspace*{-1.2cm}
\includegraphics[width=1\textwidth]{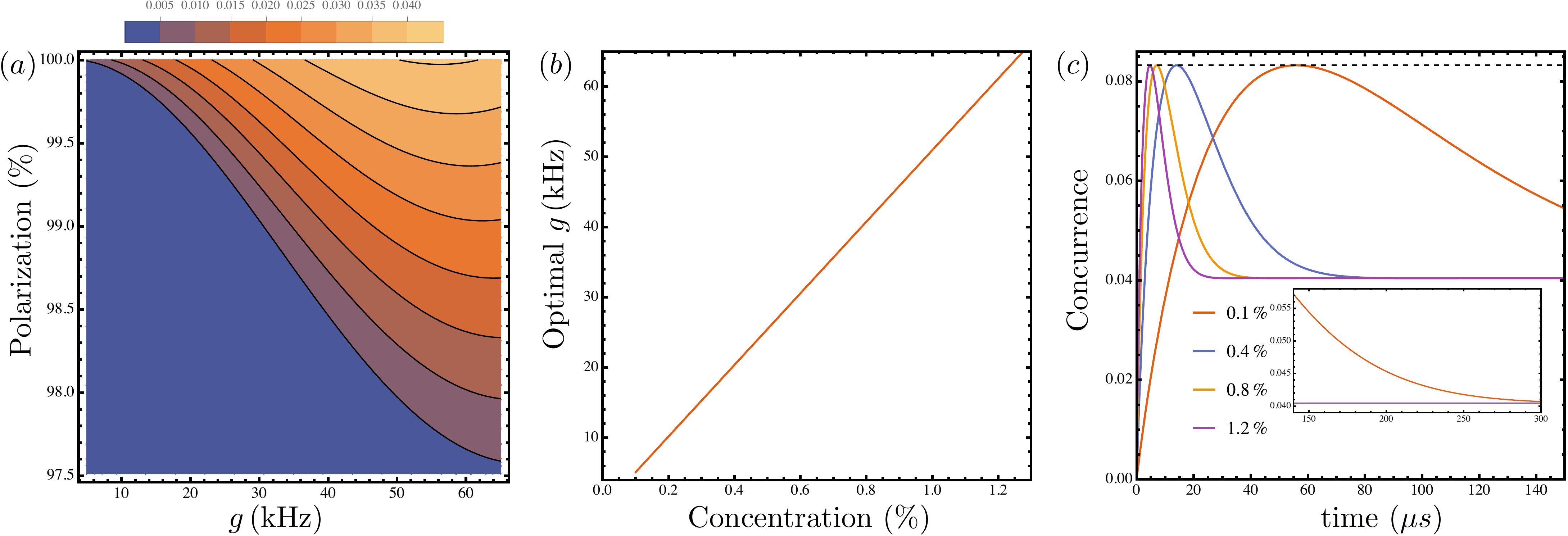}
    \caption{Entanglement predicted between the two NV centers. a) Contour plot showing the concurrence as a function of the polarization $\%$ and the inter-qubit coupling, $g$. The other parameters are $\varepsilon =1.1$ MHz, $T_h=1$ GHz and natural concentration of $^{13}$C $=1.4$ $\%$. b) The optimal values (i.e., optimized to maximize the steady state concurrence) of inter-qubit coupling for given $^{13}$C concentrations. c) Transient behaviour of concurrence for four different values of concentration and the corresponding optimal value of inter-qubit coupling. The other parameters are $\varepsilon=1.1$ MHz, $T_h=1$ GHz and 100 $\%$ polarization of \Cth constituting the cold reservoir. The initial state is chosen to be a tensor product of thermal states at the reservoir temperatures (i.e., with $n_F^c =0$ and $n_F^h\approx1/2$).The couplings are such that $g < \SI{70}{\kilo\hertz} < \Gamma^{(i)}_{1,\text{\tiny{CR}}} = \SI{250}{\kilo\hertz}$, which fits the chosen master equation regime.}
    \label{fig:conc}
\end{figure*}

\subsection{Spin diffusion}

In general, spin polarization in DNP protocols may also happen through spin diffusion, as spins interact directly through the dipole-dipole interaction Hamiltonian (similar to Eq.~\eqref{eq:dip-dip2}). % diffuse through dipole-dipole coupling. This interaction strength is typically three orders of magnitude smaller than  which is $\gamma_n$.

%nuclear spins typically interact with their environment through a three-order of magnitude lower than the gyromagnetic ratio $\gamma_n$. In particular two nuclear spins amoung the \Cth ($\beta\in\{\cth\}$)% $i^{th}$ and $j^{th}$ ,
%quantized along the same biased magnetic field,
%interact with one another through the dipole-dipole Hamiltonian of %$H_{dd}$ of Eq.~\eqref{eq:dip-dip} taking $\gamma^i\gamma^j=\gamma_n^2$.}

In a $\SI{100}{\percent}$ enriched diamond, the spin diffusion constant can be calculated directly $D_n\left(\SI{100}{\percent}\right) = \SI{6.5}{\nano\meter\squared\per\second}$. It has a square-root dependence on $n_C$ \cite{Parker2019OpticallyDiamond}. For our proposal, we estimate it to
\begin{equation}
\label{eq:Dcth}
\begin{aligned}
    D_C\left(\%\left(\cth\right)\right)=\sqrt{\frac{\%\left(\cth\right)}{\SI{100}{\percent}}}\times\SI{6.5}{\nano\meter\squared\per\second} %\SI{}{10^{-14}\centi\meter\squared\per\second}=0.65\sqrt{\frac{\rho_{^{13}C}}{\SI{1.1}{\percent}}} 
\end{aligned}
\end{equation} 

This is also consistent with the experimental value of $\SI{6.5}{\nano\meter\squared\per\second}$ obtained in \cite{Terblanche2013CDefects} for a diamond with \Cth in natural abundance. Another indirect experimental estimation gives a diffusion constant higher by one order of magnitude \cite{ajoy2019hyperpolarized}. However, this is related to a diamond being with additional nitrogen impurities which can mediate the spin diffusion. Accounting for spin diffusion, the polarization of a given bath is expected to propagate over a distance of $d=\SI{20}{\nano\meter}$ in $\approx\frac{d^2}{D_C}= \SI{1}{\minute}$. This time scale is orders of magnitude higher than the other time scales involved in this proposal which are of the order of tens of microseconds (See Fig.~\ref{fig:conc} panel c)) and previous discussions. We therefore predict this spin diffusion to be negligible in the realization of an entanglement engine. \par
In the last section, we present numerical estimations for the concurrence, a measure for entanglement. We investigate it as a function of relevant parameters of the proposal, the polarization of the nuclear spin baths, the concentration of \Cth and the coupling $g$ between the NV centers, which is determined by the distance between them. %Even in this pessimistic case, the \Cth spin diffusion appears to be negligible. 

\section{Entanglement generation}
\label{sec:ent}

Having discussed all relevant factors that determine the quantum evolution of the NV centers in our scheme, we now proceed to discuss the dynamics. For this purpose, we take the master-equation approach described in Sec. \ref{sec:intro}. %We note that in the NV setup, as we have discussed above, the pure dephasing rate and the lattice-spin relaxation rate are both negligible compared to the cross relaxation due to the \Cth environment. Therefore, we neglect them and focus only on cross relaxation. 
Analytical results on both steady-state \cite{Khandelwal2020} and transient regimes \cite{Khandelwal2021} have been previously discussed for the system. %\sk{SK: I don't understand what you mean here, Mayeul.}[]\mc{having $\Gamma_2^\alpha\ll\gamma_{1,CR}^\alpha$}.  
 Here, we focus on numerically demonstrating the generation of entanglement considering values for the parameters that correspond to the techniques and regimes described in the previous sections.
\par
In Fig. \ref{fig:conc} panel a), we show the steady-state concurrence \cite{Wootters1998} as a function of polarization and coupling between the NV centers for natural concentration of \Cth. We find that there is a large range of experimental parameters for which the qubits become entangled in the steady state. Specifically, higher the polarization (i.e., lower the effective temperature of the ``cold" reservoir), and larger the coupling between them (i.e., smaller the distance between them), the higher the steady-state entanglement. The entanglement is larger with increasing temperature gradient, but the maximum takes place at an intermediate inter-qubit coupling. The parameter range where entanglement is nonzero can be obtained exactly via the critical heat current introduced in \cite{Khandelwal2021}.
\par
In general, entanglement may appear in the transient dynamics and may disappear completely at long times. Furthermore, depending on the time-scale of the dynamics of the machine, it may be more experimentally feasible to look at short or intermediate times to detect entanglement. Importantly, the entanglement at an intermediate time may be much larger than at long times. It is therefore, also important to investigate entanglement production in the transient regime. In Fig.~\ref{fig:conc} panel c), the goal is to see how much entanglement can be created at intermediate times for different concentrations of \Cth in the diamond sample. In Fig.~\ref{fig:conc} panel b), for given concentration, we plot the optimal value of $g$ (i.e., the value which maximizes transient concurrence). In Fig.~\ref{fig:conc} panel c), we use these value of $g$ to plot the concurrence as a function of time. We find that for any given concentration, there is a $g$, such that concurrence can be maximised to the same value in the transient state, which is more than double the steady-state concurrence. Since $g$ is one of the factors that determines the time scale of the dynamics, this maximum, however, happens at different points in time. Interestingly, in the steady state, all curves approach the same value of concurrence (see inset).

\begin{table*}[h]
\centering
\begin{tabular}{@{} m{3.6cm} m{0.7cm} m{1.7cm} m{1.8cm} m{1.8cm} m{1.8cm} m{1.8cm} m{1.8cm}  @{}}

\toprule
\addlinespace[3pt]

\large{\textbf{Nuclear bath}} & &
 & \textbf{\Cth}            & \textbf{\Cth}                 & \textbf{$^1$H}    & \textbf{$^1$H}    & \textbf{$^{19}$F}  \\
 & & & $\SI{1.1}{\percent}$   & $3$ to $\SI{100}{\percent}$   & biphenyl          & Al2O3             & Al2O3 \\
\addlinespace[3pt] \hline \addlinespace[3pt] \textbf{Concentration} & $n_n$ & $\left(\SI{}{\per\nano\meter\cubed}\right)$                                
 & $1.9$                    & $5.1$ to $170$                & $41$              &                   & \\
 & $\sigma_n $ & $ \left(\SI{}{\per\nano\meter\squared}\right)$           
 &                          &                               &                   & $\gg3$             & $3$ \cite{Liu2022SurfNMR} \\
\addlinespace[3pt]                           \textbf{Polarization radius} & $R_M $ & $\left(\SI{}{\nano\meter}\right)$                             
 & $1.33$                   & $1.33$                        &                   &                   &  \\
\addlinespace[3pt]                           \textbf{NV center depth} & $d_{\text{\tiny{NV}}}$ & $ \left(\SI{}{\nano\meter}\right)$                             
 &                          &                               & $6$               & $10$              & $10$ \\
\addlinespace[3pt]                          \textbf{Number of nuclei} & &
 & $20$                     & $20 \times \dfrac{n_c}{n_c^0}$ &                   &                   & \\
\addlinespace[3pt] \hline \addlinespace[3pt] \textbf{Polarization Rate} & $R$ & $ \left(\SI{}{\hertz}\right)$                            
 &$9000$                    & $9000$                        & $7500$            & $375$             & $154$ \\
\addlinespace[3pt]                           \textbf{Maximal polarization} &  $p_{\text{max}}$ & $ \left(\SI{}{\percent}\right)$                       
 & $>99$                    & $>99$                         & few $\SI{}{\percent}$                  & $\SI{}{\percent}$              & $0.3$ \\
 \addlinespace[3pt]                         \textbf{Diffusion constant}   & $D$ & $\left(\SI{}{\nano\meter\squared\per\second}\right)$            
 & $7$ \cite{ajoy2019hyperpolarized}         & Eq.~\eqref{eq:Dcth} &571               & 700               & 700 \\
\addlinespace[3pt] \hline \addlinespace[3pt] \textbf{Gyromagnetic ratio} & $\gamma_n$ & $ \left(\SI{}{\mega\hertz\per\tesla}\right)$                
 & $10.7$                   & $10.7$                        & $42.6$            & $42.6$            & $40.0$\\
\addlinespace[3pt]                           \textbf{Qubit energy}      & $\varepsilon_{\text{\tiny{L,R}}}$ & $ \left(\SI{}{\mega\hertz}\right)$                  
 & $1.1$                    & $1.1$                         & $4.4$             & $4.4$             & $4.1$ \\
\addlinespace[3pt]                            \textbf{Qubit/bath coupling}  & $\Gamma^{(i)}_{1,\text{\tiny{CR}}}$ & $ \left(\SI{}{\kilo\hertz}\right)$                  
 & $250$ & Eq.~\eqref{eq:gcrcth} & few $1000$ & few $1000$ & few $1000$ \\
\addlinespace[3pt]                           \textbf{Decoherence}    & $\Gamma^{(i)}_2$ & $ \left(\SI{}{\kilo\hertz}\right)$                        
 & $1.5$ \cite{Mizuochi2009} & Eq.~\eqref{eq:g2cth}                              &   ($4.5$-$14$)$10^3$               & few ${100}$  &  \\
\addlinespace[3pt]                           \textbf{Spin-latice relaxation} & $\Gamma^{(i)}_{1,\text{\tiny{SL}}}$ & $ \left(\SI{}{\hertz}\right)$                           
 & $150$ & $150$ & $150$ & $150$ & $150$ \\
\addlinespace[3pt] \hline \addlinespace[3pt] \textbf{Reference}   &   &  
& \cite{broadway2018quantumHyperpol,Rizzato2022}                          &                               & \cite{Healey2021} & \cite{Rizzato2022} & \cite{Rizzato2022}\\   

\bottomrule
\hline
\end{tabular}
\caption{Dynamical Nuclear Polarization parameters for different nuclear spins and concentration. Missing values are either not applicable or not available from the literature}
\label{tab:1}
\end{table*}

% \sk{[SK: this is not what is shown in this figure.]In Fig. \ref{fig:transient1}, we show concurrence as a function of time for varying coupling rates. The behaviour is the opposite when $g$ or $\Gamma^{\alpha}_1$ and $\Gamma^{\alpha}_2$ are changed. As $g$ is increases, the peak concurrence increases. However, as $\Gamma^{\alpha}_1=\Gamma^{\alpha}_2$ is increased, the peak concurrence decreases. As expected, dephasing is detrimental to entanglement production in the machine.}

%\begin{figure}
%  \begin{subfigure}{6cm}
%    \centering\includegraphics[scale = 0.5]{qm1.png}
%    \caption{plots of time independent and fin dependent solution at different times, the number of spins polarized are per unit volume in SI units, i.e., $m^{-3}$}  \end{subfigure}
%  \begin{subfigure}{6cm}
%    \centering\includegraphics[scale = 0.5]{qm2.png}
%    \caption{Plots of time dependent temperature at different times and distances from the center of polarization.}
%  \end{subfigure}
%\end{figure}

%\begin{figure*}
%\centering
%\includegraphics[scale = 0.5]{qm1.png}
%\caption{plots of time independent and fin dependent solution at different times, the number of spins polarized are per unit volume in SI units, i.e., $m^{-3}$}
%\label{fig:tdep}
%\end{figure*}

%\begin{figure*}
%\centering
%\includegraphics[scale = 0.5]{qm2.png}
%\caption{Plots of time dependent temperature at different times and distances from the center of polarization.}
%\label{fig:tempplots}
%\end{figure*}

\section{Conclusion}\label{sec:diss}

In this work, we have proposed an experimental setup to realize a two-qubit entanglement engine on an NV-center platform. We discuss the implementation of two interacting qubits, as well as the realization of two independent nuclear spin baths characterized by different temperatures. This temperature gradient is realized effectively through a polarization gradient created using DNP. It is interesting to note that certain DNP protocols can also engineer effective baths at negative temperature~\cite{Healey2021,Rizzato2022}, corresponding to spin bath polarized in their ``up" state. From a thermodynamic point of view, this constitutes an interesting resource, already discussed in few works in the context of quantum thermal machines~\cite{Assis2019, Nettersheim2022, brask2022operational}.

A key requirement of a two-qubit engine is to maintain a long lasting out-of-equilibrium situation. In schemes involving local interactions, this appears particularly difficult. It is indeed required to avoid any direct coupling between the two thermal reservoirs, at a similar distance to one another than the one between qubits, while keeping dominating qubits/qubit and qubit/bath couplings with similar order of magnitude. In our scheme, this is made possible by exploiting the large difference between the gyromagnetic moments of, on one side the nuclear spins for the baths and on the other side the electronic spins for the two interacting qubits. For instance, in the alternative of considering baths made of electronic spins, all interaction strengths would be set by the gyromagnetic moments of electronic spins, preventing the realization of two independent reservoirs interacting distinctively with each NV center. 

As a possible alternative, one may consider different nuclear spins. For instance, in references~\cite{Healey2021,Rizzato2022}, local efficient polarization of surface nuclear spins \Fnine or \Hone has been demonstrated. The relevant references and parameters to investigate such configurations together with alternative \Cth volume concentrations are provided in Table \ref{tab:1}. In those cases, one could take profit of the potential shorter distance between nuclei to strengthen the thermal reservoir approximation by involving more of them in each bath enabling interactions between them. The qubit/bath coupling could then be set by the NV centers' depths. In such cases however, the spin diffusion to be considered may then not be negligible anymore and would required a proper modeling. Such alternatives may constitute interesting schemes to be investigated both theoretically and experimentally.

\section*{Acknowledgements} 
The authors thank Christophe Galland and Jonatan Bohr Brask for discussions at an early stage of this work. All authors acknowledge support from the Swiss National Science Foundation; GH and SKh through
the SNSF starting grant PRIMA PR00P2\_179748  and MC and SKu through Ambizione grant No. 185824.

\bibliography{References}

\end{document}